\documentclass{elsart}
\usepackage{amsmath,amssymb,epsfig}

\begin{document}

\begin{frontmatter}

\title{Effects of decision-making on the transport costs across complex networks}
\author{Sean Gourley},  
\ead{sean.gourley@balliol.ox.ac.uk}
\author{Neil. F. Johnson}
\ead{n.johnson@physics.ox.ac.uk}
 \thanks[label1]{We thank EPSRC (U.K.) for funding though grant EP/D001382, and the  
European Union for funding under the MMCOMNET programme.}
\address{Physics Department, Clarendon Laboratory, Parks Road, Oxford, OX1 3PU, U.K. }
\begin{abstract}
We analyse the effects of agents' decisions on the creation of congestion on a centralised network with ring-and-hub topology. We show that there are two classes of agents each displaying a distinct set of behaviours. The dynamics of the system are driven by an interplay between the formation of, and transition between, unique stable states that arise as the network is varied. We show how the flow of objects across the network can be understood in terms of the ordering and allocation of strategies. Our results show that the existence of congestion in a network is a dynamic process that is as much dependent on the agents' decisions as it is on the structure of the network itself. 
\end{abstract}
\end{frontmatter}
\section{Introduction}
Understanding how motorists' individual decisions affect the traffic patterns which emerge on road networks, is of great practical importance \cite{1}. It also represents a fascinating theoretical problem from the point of view of transport on networks. Indeed, the study of the functional properties of networks is gaining increased attention across a range of disciplines  \cite{1,2,3,4}.  Of particular interest is the fact that congestion at various critical points on the network can dramatically reduce the efficiency of the network. Ashton $et~al$ have presented an exactly solvable model of a ring-and-hub network  \cite{5} which extended the model of Ref. \cite{6} to include congestion costs on the central hub or hubs. 

The quickest route across the network is easy to determine when you are the only agent on the network. However this is rarely the case in real-world problems, where you have multiple agents all trying to minimise the time/cost of traversing the network. When this happens you see congestion at the major shortcut points, e.g. the random links in small world networks, or at the major hubs in a scale free network. Once the connection becomes congested it no longer represents the shortest pathway across the network. Agents observe this and modify their behaviour, often creating a new congestion points elsewhere on the network. Congestion arises then not solely as a result of the network topology, rather it occurs as a result of the dynamic interplay between the structure of the network and the decisions of the agents using it. In this paper we introduce a model for describing the effects of agents' decisions on the creation of congestion within a ring-and-hub topology, this work extends the model introduced by Ashton $et~al$ by introducing {\em decision-making agents} onto the network. Agents use their own strategies to make inductive decisions about the future behaviour of the system in order to find the cheapest pathway across it. Our model lends itself to real life problems such as communication across social networks, flow of data across the internet, traffic flow, or any situation where competing agents have to navigate a network where congestion is a factor. This model is constructed in an attempt to understand how agents on centralised networks respond to a dynamic pricing structure, and the effect that these charges have on congestion.

\section{System Set-up}
The simulation consists of $N$ agents and a central hub of capacity $L$. Each agent is connected to their nearest neighbours by an undirected link of unit length. These links form a peripheral pathway around the outside of the network. Each agent also have the possibility of being connected to another point on the network through the central hub. If this pathway exists it is known as the hub pathway and the number of these in the network is defined to be  $\lambda$. Through these sets of connections the agents form a combined ring-and-hub topology, i.e. a hub and spoke network. Each agent $A(i)$ must transport himself (e.g. a car containing himself, or a message) from one location on the network to a randomly selected final destination $A(j)$ at another point on the network. If the agent $A(i)$ is connected to the central hub they have the option of using this resource with an associated cost $C_{central}$, or they can use the peripheral pathway constructed from connections between nearest neighbours at a cost of $C_{out}$. The goal of each agent is to minimise the cost of transporting the object/data to its final destination. 

\begin{figure}[h]
\begin{center}
\includegraphics[width=0.35\textwidth]{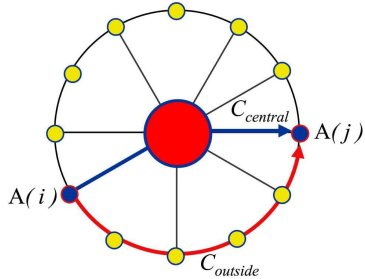}
\end{center}
\caption{ Our model network with the nodes connected to nearest neighbours around 
the outside and the central hub located in the middle. Agent $A(i)$ is randomly connected to another point on the network $A(j)$ and the costs of the transport are shown as $C_{central}$, $C_{outside}$.}
\label{fig:1}
\end{figure}

There are costs associated with each decision and these are listed below in equation (1), where the cost of using the central hub is a variable cost that is dependent both on the actions of the agents within the group and the capacity of the hub. The central hub has a finite capacity given by $L$, if this capacity is reached then the hub is congested and a congestion charge $cc$ (time/money) is imposed on all traffic through the hub. There are many ways to implement a congestion charge depending on the system under investigation, but for the purposes of this paper we will choose a digital cost structure (as shown in figure 1b), where each connection to the hub is $\frac{1}{2}$ a unit length and the congestion charge only applies when $>L$ agents use the central hub. We select a digital price structure for the congestion charge for two reasons, firstly when designing systems with a dynamic pricing component, a digital structure is far easier to implement than an analogue one, as it is a simple on/off switch like the congestion charge in central London. Secondly, a digital structure captures the elements of many real world congestion problems, such as the movement of road traffic, which undergoes an abrupt transition from free flowing to stop-go traffic at a critical density \cite{7}.

In contrast to the variable pricing structure of the central hub, the cost of using the peripheral pathway is determined only by the number of nodes traversed and as such is a fixed cost. There is a cost of  $\beta=1$ associated with travelling between two neighbouring nodes on the network. The transport costs across the network are then given by;

\begin{eqnarray}
C_{central} &=& \left\{ \begin{array}{ll}
2 \left(\frac{\beta}{2}\right) & \textrm{if}~ N_{central}\leq L \\
cc + \beta & \textrm{if}~N_{central}> L\end{array} \right. \nonumber\\ 
C_{out}&=&n\beta
\end{eqnarray}
\begin{figure}[h]
\begin{center}
\includegraphics[width=0.7\textwidth]{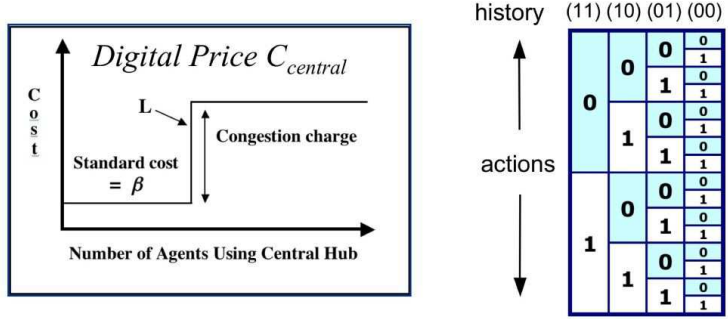}
\end{center}
\caption{ (a) Digital pricing structure for the central hub with capacity $L$. 
A standard cost of $\beta=1$ is applied before the capacity reached. After 
capacity is reached the cost of using the hub increases by fixed amount 
$cc$.
(b) The $m=2$ set of strategies allocated to agents in this simulation. 
Each column represents a different realisation of the history string, 
with rows containing the actions. Action 1 corresponds to using the central hub, and action 
0 is to take the peripheral path.}
\label{fig:2}
\end{figure}
Where $n$ is the number of nodes traversed, $cc$ is the congestion charge, which can vary between $0$ and $N/2$, and  $\beta$ is the standard or unit cost. The connections between nodes on the network are undirected links of unit length; hence data can travel either way round the network to its destination. The maximum distance an object can travel is then $N/2$. The connections to the central hub are directed and unique to the agent, hence only agents directly connected to the hub are able to use it. If an agent is not connected to the central hub, they are forced to use a peripheral pathway.

In order to determine the quickest/cheapest route, each agent is randomly assigned $s=2$ strategies from a pool of binary strategies. For $m=2$ these strategies take the form $(1011)$, where each digit in the strategy sequence corresponds to an action associated with the history string of the same position $(11,10,01,00)$ i.e. for history string $(10)$ we have action $0$. Here action $1$ denotes a decision to use the central hub, whilst $0$ corresponds to not using the central hub. The strategy table (shown in figure 2) is self-similar in nature, and for every node visited by the global history string, another column of the table is accessed to reveal differences in strategies. 

At each time-step in the game every agent with a connection to the central hub must make a decision whether or not to use the central hub, the decision can be summarized as ``through the middle, or around the outside?" Their decision is dictated by the relative success of the two strategies that they hold. The agents make their decision based on the action associated with their highest scoring strategy, and if the two strategies are tied then the agent will flip a coin to decide. If the agent chooses action 1 and $C_{central} < C_{out}$ then the agent has made the correct decision and the success of their strategy will be reinforced with an increase in it's virtual score of +1 points, else if $C_{out} > C_{central}$ then the strategy will be penalised by -1 points. The reverse applies if the agent's high scoring strategy predicts action 0. At time $t=t+1$, using the newly updated strategies, the agent again makes a decision about the cheapest pathway to use, and the above process is repeated. 

\section{Results and Analysis for Variable Network}

At the start of the simulation there are no connections to the central hub and each agent is only connected to their nearest neighbours. At each time-step a randomly chosen agent is connected to the central hub such that $\lambda =\lambda+1$. With this new network in place, the strategies and destinations are then reassigned amongst the agents and the simulation is repeated with the agents competing to minimise their transport costs across the new network. This process is repeated until all agents have a connection to the central hub. We have run this simulation with $N=101$ agents, a memory length of $m=2$ and $s=2$ strategies assigned per agent. The central hub has a capacity given by $L$ and a congestion charge determined by the digital price structure shown above in figure 2a. The simulation is run for 10,000 time-steps which constitutes one run, with each value of $\lambda$ representing the average of 1000 runs.

The global cost per agent of transportation across the network is defined as $g(\lambda)$, and is displayed in figure 3a for various values of $cc$ in a network with $L=40$. The transport cost, is initially the same for all values of $cc$ and starts at $g(0)\sim25$. This value is then the cost of transporting data/objects across the network with no connections to the central hub, and is given by $g(0) = \frac{N}{4}\beta$. As $\lambda$ increases, $g$ decreases linearly for all values of $cc$ up to a critical point at  $\lambda \sim 50$. The curves for the various values of $cc$ then diverge and follow distinct pathways. We can divide the plots into two groups; high penalty $cc>25$ and low penalty $cc<25$. For the high penalty group $g(\lambda)$ increases rapidly after the critical point until the emergence of a stable state at $\lambda \sim 65$ where $g$ stays relatively constant before increasing further as  $\lambda$ tends to $N$. For the low penalty group an increase in $g(\lambda)$  after the critical point is also observed, however after this initial increase the cost of transportation across the network again begins to fall. 

\begin{figure}[h]
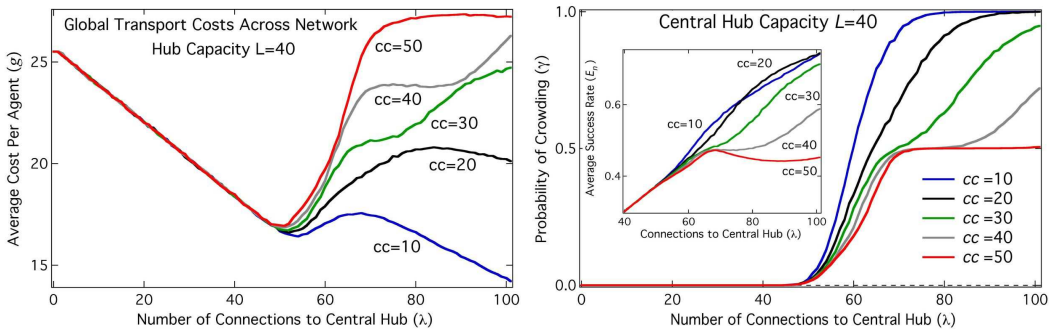

\includegraphics[width=0.5\textwidth]{figure3a.epsf}
\includegraphics[width=0.5\textwidth]{figure3b.epsf}
\caption{(a) This graph displays the results from the simulation showing the global 
transport cost across the network g($\lambda$) as a function of the number 
of agents connected to the central hub. 
(b) This graph shows the probability of the central hub being crowded 
($\gamma$). Inset shows the agents success rate in predicting the 
cheapest pathway across the network.}
\label{fig:3}
\end{figure}

In figure 3b we see the plot of the probability of the central hub being crowded, $\gamma(\lambda)$. For values of $\lambda<50$, $\gamma=0$ as the central hub is never overcrowded. For the low penalty systems little or no time is spent in the   $\gamma=0.5$ state. Whereas for high penalty systems as $cc$ increases, the amount of time spent in this state also increases, and when $cc=50$ the system does not leave the  $\gamma=0.5$ state. The inset of figure 3b shows the agents average success rate in predicting the correct transport pathway across the network, which we define to be $E(\lambda)$. 

We can divide the agents into two groups based on the distance that they have to travel and the size of the congestion charge, where $N_{short}$ denotes the average number of `short trip' agents and $N_{long}$ the average number of `long trip' agents. $N_{long}$ agents have to travel a distance that is greater than $cc$. Hence it is always advantageous for the long trip agents to use the central hub irrespective of the actions of other agents. The short trip agents have a distance to travel which is less than the size of the congestion charge. Hence the short trip agents will want to use the central hub provided it is $not$ congested. However if the central hub is congested, then it will prove cheaper to use the peripheral pathway. Thus for the $N_{short}$ agents, the correct decision is dependent on the collective actions of the group. Because the agents' final destinations are distributed randomly, we get for $cc<N/2$;

\begin{eqnarray}
N_{short} &=& \left(\frac{\textrm{cost of using crowded hub}}{\textrm{cost of maximum path across network}}\right)N_{total} + \beta \\
&=&2 cc +1 \\
\nonumber\\
N_{long} &=& N_{total} - N_{short} \\
& = &N_{total} - 2cc-1
\end{eqnarray}

If $cc<N/2$ then there are no long trip agents and $N_{short}=N_{total}$. Because the size of the congestion charge remains constant, the sizes of these two groups stays fixed throughout the duration of the game. There are then two separate history stings associated with the game, $\mu_{short}$ and  $\mu_{long}$, one for each group of agents. The agents in each group can then effectively be treated as a cohesive unit, whose actions are jointly determined by their respective values of  $\mu$ and the initial strategy distribution amongst the agents within the group. 

The game starts with $\lambda=0$ and the hub is under-subscribed with no congestion, this state will be called State I. The history string for both groups of agents is then $\mu_{long}=\mu_{short}=(0000...)$, where $0$ denotes the global un-crowded result. This history string only visits one node on the De Bruijn graph (00) (figure 4a) and as such results in a compression of the strategy space from the original pool of 16, to just two. The two strategies are then (0$\mid$1) and (0$\mid$0), where the first term denotes the history and the second the agents' action. Because the history string consists of consecutive 0's the virtual point scores for the two strategies diverge linearly over time as shown in figure 4b. This gives rise to the state level diagram shown in figure 4c with two states corresponding to the two different strategies (0$\mid$1) and (0$\mid$0). Because each agent is randomly assigned $s=2$ strategies, and plays the higher scoring of the two, the populations of the levels is $3N/4$ and $N/4$ respectively. Thus for every link that is added to the central hub, the usage of the central hub ($N(1)$) increases on average by $3/4$ agents, giving us for State I;
\begin{eqnarray}
g(\lambda) & = & (\textrm{average peripheral cost})-\lambda(\textrm{$N_{central}/N_{total}$}) C_{central} \nonumber\\
& =& \frac{N}{4} - \frac{3 \lambda}{4} \beta
\end{eqnarray}
This process continues as long as the system remains in the un-crowded state with   $N(1) <L$. We can then define a critical point as the point where the system has a $50\%$ chance of $N(1)>L$. Thus the critical point occurs at $\lambda =(4/3)L$, we have $N(1)=L$ and the system moves out of State I and into State II. The change in state corresponds to a change in history string for short trip agents, with $\mu_{short} = (0001...)$. However the history string for the long trip agents remains fixed at  $\mu_{long} = (0000...)$.

\begin{figure}[h]
\begin{center}
\includegraphics[width=0.9\textwidth]{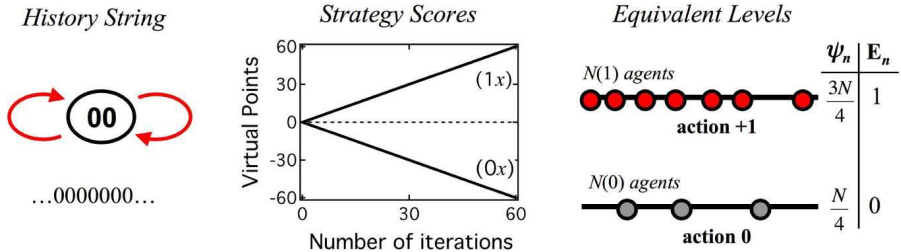}
\end{center}
\caption{(a) The path across the de Bruijn graph for State 1. (b) The 
corresponding virtual points scores for the two classes of strategies 
used by agents to make their decisions, (c) Equivalent levels, along 
with associated populations and success rates.}
\label{fig:4}
\end{figure}

In State II (see figure 5a) the system visits two extra nodes (01) and (10). Visiting the nodes has two effects, the first is that for part of the time the central hub is crowded and a congestion charge is applied to all agents who use it. This crowding is responsible for the increase in $g(\lambda)$   that occurs after the critical point. The increase in $g(\lambda)$ is a gradual one since the probability of residing in the State II is governed by a binomial distribution. The second effect of the nodes is to increase the number of strategies available to the agents, which in turn changes the number of bands in the state level diagram. These two nodes can be thought of as providing extra degrees of freedom for the system and as such resolves the $(1xx)$ strategy into four new strategies (111, 110, 101, 100). This extra resolution means that for node (10) a percentage of the agents that were taking action 1 before the transition are now taking action 0, and as such $N(1)$ must initially be less that $L$ in State II. The same strategy splitting effect occurs for the $(0xx)$ group of strategies, which if $s=1$ would cancel the above effects and the system would move out of this state. However because $s=2$ the distribution of strategies is top heavy for the strategies with $(00\mid1)$, the splitting of strategies instead acts to reduce $N(1)$ and a buffer is created.

\begin{figure}[h]
\begin{center}
\includegraphics[width=.99\textwidth]{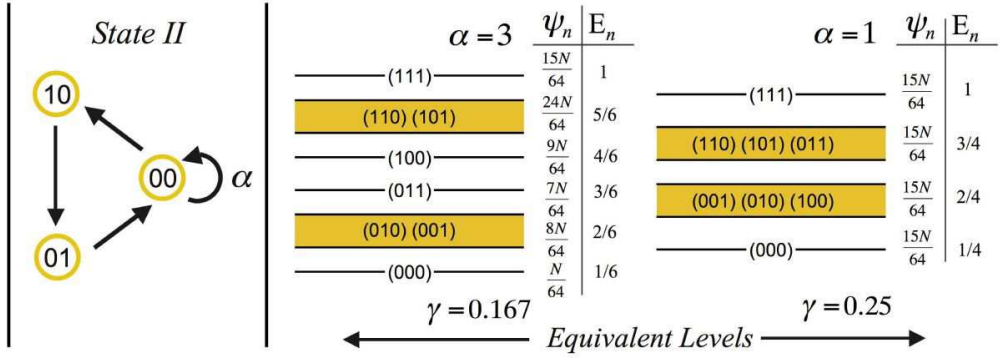}
\end{center}
\caption{(a) de Bruijn graph for State II, system visits three nodes and $\alpha$ is the number of times the system returns to (00) node. (b) State level diagram for State II, the wide bands represent groups of strategies with the same mean success rate.}
\label{fig:5}
\end{figure}

In State II, the system traverses the de Bruijn graph in a cyclic fashion, returning to the (00) node each time (figure 5a). However after taking this pathway the original ordering of the strategies is different. The system then resets itself by returning to the (00) node $\alpha$  times. As $\lambda$ increases the value of  $\alpha$ decreases, from $\alpha=3$ initially to $\alpha =1$ immediately before leaving State II. This change in  $\alpha$  reduces the number of unique levels in the state diagram, and alters the make-up of the strategies within them (figure 5b).

The wider bands in the state level diagram represent sets of strategies which have the same average success rate over time $(E_n)$, but vary about $E_n$ during the cycle around the graph. The strategies will vary about this mean, but on $\frac{2}{3}$ of the nodes they will have equal virtual point scores. On these nodes agents that hold two strategies from within the same band are forced to flip a coin in order to decide which strategy to play. If the agent's two tied strategies make the same predictions for the particular node, then it does not increase the disorder in the system. However if the two strategies make different predictions, flipping the coin does influence their action and this agent falls into a new group of agents which we shall call the undecided group denoted by $N(\frac{1}{2})$. There are thus three elements in the system, $N(1)$ the number of agents choosing action 1 with certainty, $N(0)$ the number of agents choosing action 0 with certainty and $N(\frac{1}{2})$ the number of agents whose decision is determined by chance. Because the long trip agents effectively only have two non-equal strategies to play, the $N(\frac{1}{2})$ agents come exclusively from the short trip population. The relative sizes of these three groups determines which state the system will reside in and hence the probability that the central hub is crowded. 

\begin{figure}[h]
\begin{center}
\includegraphics[width=0.9\textwidth]{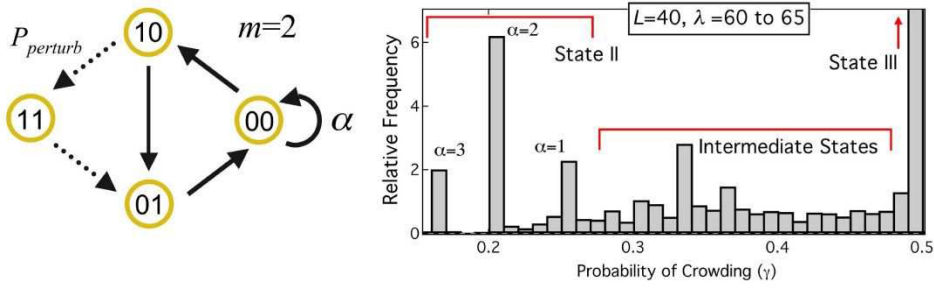}
\end{center}
\caption{((a) de Bruijn graph for the intermediate state between States II and III. 
Probability of visiting the (11) node given by 
$P_{perturb}$. (b) The histogram reveals the existence of intermediate states of $\gamma$ for $60<\lambda<65$.}
\label{fig:6}
\end{figure}
A global `1' result on the (10) node will shift the system out of State II and into State III. In order to determine the probability of this occurring we need to consider the size of the $N(1)$ and $N(\frac{1}{2})$  components for this state. We will look specifically at the $\alpha =1$ distribution since this is the last cycle the system visits before moving into State III. Analysis of  $\psi_n$ and the band levels in figure 5c gives us, $N\left(\frac{1}{2}\right) = \frac{8N}{64}$. If $N(\frac{1}{2})$ is greater than the difference between the resource level and $N(1)$, then the system will be randomly 'kicked out' of the stable cycle in State II and will briefly reside in State III before returning. This process is shown in figure 6a. If we define the buffer as, $\Delta = L - N(1) - N\left(\frac{1}{2}\right)$. We can then use binomial probability distribution to determine the likelihood that the system will be 'randomly' perturbed into a new state. Doing this gives us;
\begin{displaymath}
P_{peturb} = \sum_{k = \Delta}^{ N\left(\frac{1}{2}\right)} P(k\mid N\left(\frac{1}{2}\right)) = \sum_{k = \Delta}^{ N\left(\frac{1}{2}\right)}
\frac{N\left(\frac{1}{2}\right)!}{k!( N\left(\frac{1}{2}\right)    -k)!}
\left(\frac{1}{2}\right)^k \left(1-\frac{1}{2}\right)^{N\left(\frac{1}{2}\right)-k}
\end{displaymath}

The effect of the $N(\frac{1}{2})$ agents is to act as noise which can randomly perturb the system into intermediate states between States II and III, these can be seen clearly in figure 6b for  $0.25<\gamma< 0.5$.

Each new state that the system visits has a unique cost, and gamma associated with it. These can be determined by careful analysis of the state level diagrams, which gives us a general expression for the global transport cost;
\begin{eqnarray}
g(\lambda) &=& \sum_{n=1}^i g(\lambda)_n P(\lambda)_n 
\end{eqnarray}

where $g(\lambda)_n$ is the cost associated with State $n$, and $P(\lambda)_n$ is the probability of residing in this state. Elsewhere we will present a detailed analysis of these states and the transitions between them. Including a discussion of the high $\gamma$ states which occur in the remaining part of the parameter space.

\section{Summary}

This paper has shown that the agents in the network can be divided into two groups, where the size of each group is determined by the maximum path length across the network and $cc$. The addition of new links to the network brings about two main processes that serve to drive the system into new states. The first of these processes is the addition of $N(1)$ agents who use the central hub with $certainty$. The second process is a more unpredictable one with the addition of `random' $N(\frac{1}{2})$ agents to the system. If $N(1)+N(\frac{1}{2})>L$, there is a finite probability that the $N(\frac{1}{2})$ agents will randomly perturb the system into new states, but these states are not permanent. However for any node on the graph if $N(1)>L$, the change is permanent and a new state is formed that has a unique and `predictable' cycle associated with it. This new state and the associated cycle around the de Bruijn graph has the effect, when compared with the previous state, of changing the history strings for the $N_{short}$ agents, which in turn re-orders the agents strategies. This re-ordering of strategies reduces the number of short trip agents using the central hub, and in effect acts as a buffer to produce stable states in the system. Thus the system moves between different states as the connections to the central hub are varied, each of these states has a unique behaviour associated with it and some states may be more desirable than others depending on the goal of the system designer. Our results show that whilst some states remain robust to external influences, it is possible to make large scale changes in efficiency with minor adjustments to network structure or pricing.

\end{document}